\documentclass{camera}
\usepackage{epsfig}
\begin{document}
\begin{flushright}
FERMILAB-CONF-01/092-E\footnote{Invited Talk to appear in
``Proceedings of the LEPTre Meeting on LEP Physics'', Rome April 2001}
\end{flushright}
%
\title{HIGGS DISCOVERY BEFORE LHC ?}

%
\author{Giorgio Chiarelli \\
for the CDF and D0 Collaborations }

%
\organization{I.N.F.N. Sezione di Pisa}

\maketitle

%

\section{Introduction}
The standard model (SM) of fundamental interactions, has been the
successful theory over 
the last 25 years. 
The overall success of the SM in describing the elementary
interactions, the discovery of gauge bosons at CERN in the eighties as 
well as the top discovery at Fermilab in 1995, strengthened the
expectation that the Higgs mechanism is the one that gives mass to
all particles. 
At the moment the Higgs particle is the only missing pieces of the puzzle.

The sensitivity of parameters of the electroweak theory to the mass of the top quark
and of the W boson has been exploited to provide limits on the mass of the 
Higgs particle ($M_H$). Due to the logarithmic dependence of $M_H$ to the
ratio of
$M_W/M_{top}$, a small change in the central values translates into
a large change in
the limit on $M_H$. At present (Spring 2001), the
current 95\% CL lower bound is 212 GeV/c$^2$ while the upper limit from LEP
experiments is 113.5 GeV/c$^2$ with the additional hint of a possible
signal at 115 GeV/c$^2$\cite{lathuile}.
Due to its coupling Higgs decays into the heaviest possible pair of
particles, therefore for $M_{H}$ below 130 GeV/c$^2$ (low mass region) the
most important channels are $b$ or $\tau$ pairs, while for heavier masses,
the branching fraction into vector boson pairs becomes dominant.
A hadron collider provides excellent
chances to discover the Higgs given that (in the low mass region) 
tagging of b-jets would be available.
The tool became a reality at the Tevatron during the search for top and is
now taken for granted in any experiment at hadron colliders.

\begin{figure}[htb]
\centering
\epsfig{figure=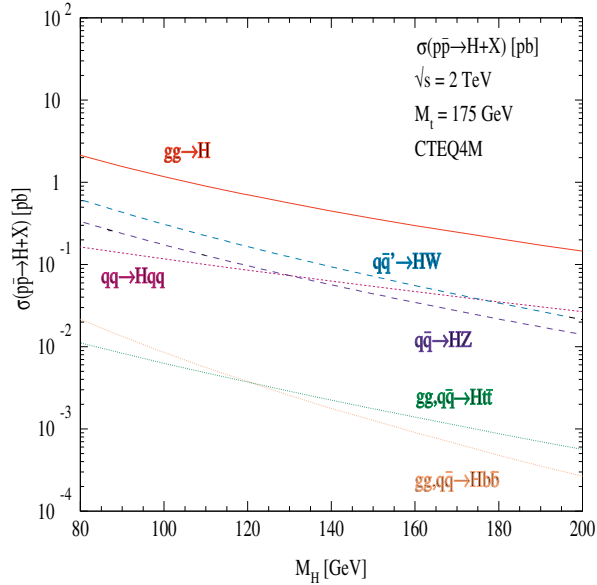,width=.51\linewidth}
\caption{Higgs Production Cross Section}
\label{fig:xsec}
\end{figure}

 In fig.~\ref{fig:xsec} Higgs production cross section at 2 TeV is shown. 
While the gluon fusion mechanism is by large the dominating generating process,
the signal over background ratio is such that at low mass the
associated production of Higgs particle and vector boson (W,Z) is
better suited for a discovery. At the Tevatron the interaction cross
section is about 70 mb. About 70\% of it provides events that are
visible in the experiments while the Higgs cross section is in the order
of 1 pb or less. Therefore an efficient trigger must operate to provide a
sample where to look for the needle in the haystack (1 in 10$^{10}$
events). 

In Run I CDF and D0 searched for Higgs produced in association with W and Z by
 exploiting samples of events collected with non-dedicated triggers. 
In specific both 
experiments looked in the "high $P_T$ lepton" and in the "missing $E_T$"
samples with the
additional selection of two b-tagged jets. Despite the limited sensitivity
to SM Higgs, the D0 and CDF analyses were useful to develop tools, study
detector efficiencies and background.
Those analyses also set the stage for studies dedicated to better 
understanding of the physics reach of CDF and D0 in Run II and demonstrated the
key role assumed by b-tagging techniques in this search.

\section{Run II Studies}
\begin{figure}[htb]
\centering
\epsfig{figure=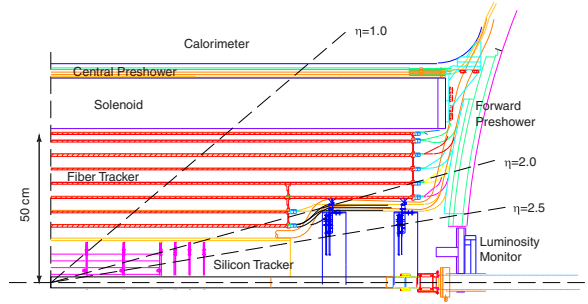,width=.5\linewidth}
\caption{Cross sectional view of the D0 Upgrade}
\label{fig:dzu}
\end{figure}

The success of Run I 
paved the way to the upgrade of the Tevatron complex. This in turn
lead to the
upgrades of the CDF and D0 detectors.
While the design energy (2 TeV in the center of mass) was not reached
in Run I, the design instantaneous luminosity
($10^{30} cm^{-2} s^{-1}$) was routinely exceeded during the 1992-1995
data taking period. The experience gained
 during the high luminosity running led to an upgrade of the machine in which more
 luminosity was obtained by introducing more bunches (36x36 in Run IIa up 
from 6x6
in Run I). In this way the luminosity is increased while the average
number of 
interactions per crossing is kept low. The price paid was the complete rebuilding of 
the front-end electronics to match the new interbunch 
(from 3.5 $\mu$s down to 396 ns in Run IIa and
 then 132 ns in Run IIb). At the same time CDF and D0 rebuilt their
tracking systems. 
D0 added a magnetic field (2 T) and replaced the older tracking with a
Fiber Tracker
 supplemented by a large silicon vertex detector (fig.~\ref{fig:dzu}). 
CDF replaced the old
 gas-based tracking chamber with a new drift chamber (the COT) and completely
 rebuilt the silicon system, which is now made of 7 layers of double sided silicon 
detectors, covering up to $|\eta|<2$. It provides a standalone system with secondary vertex
 recognition capability. Furthermore CDF and D0 largely improved their triggering
 capabilities and are now able to identify high momentum tracks at Level 1
(i.e. between crossings) and identify tracks
 displaced by the primary vertex at Level 2. The Secondary
Vertex Tracker (SVT) is operational at CDF and the D0 Level 2 displaced
vertex trigger
(STT) will be operational next year~\cite{cdftdr,d0tdr}. Both experiments
also extended their
capability to trigger on electrons and muons. The
overall goal is to retain
and exceed Run I physics capabilities through Run II.

CDF and D0 set forth a combined effort to understand the physics reach for Higgs
searches of the two detectors\cite{higgswg}. In order to do so two
complementary paths were
followed: in one simple rescaling of results obtained in Run I was
used,
while at the same time a parametric Monte Carlo (based on performances
averaged
between CDF and D0) was tuned to reproduce Run I results. In this way we were
able to estimate physics capabilities using backgrounds and
efficiencies tuned on data. Essentially all mass spectrum was studied, although most
of the efforts concentrated on the "low mass" region.
In this region the most promising channel is the associated production of
Higgs and vector boson (W or Z), with Higgs decaying into $b \overline{b}$
pairs. In
the high mass region, where the Higgs decays into vector boson pairs,
the events have a clean signature and the Higgs generation
through gluon fusion becomes the most important production process.
The results shown here are obtained by using the parametric Monte
Carlo. This allows keeping into
account the larger acceptances of the tracking system (and therefore
the improved b-tagging capabilities) as well as improvements in the 
algorithms and in trigger capabilities. As the Higgs particle should
also appear as a bump in the invariant mass of two b-jets, a lot of
effort was devoted to an improvement of the jet energy resolution.
In Run I, by using dedicated
corrections, CDF was able to reduce the
invariant mass resolution in its $Z\rightarrow b\overline{b}$ sample
to 12\%.
Different studies shown that a 10\% resolution is achievable.
CDF plans to exploit the SVT to select a
sample of $Z\rightarrow b \overline{b}$ events. D0
has similar plans based on its STT. This
trigger will of course also select events containing Higgs~\cite{ciro}. 
The channels we present here are the low mass $WH$ and $ZH$, with
$H\rightarrow b\overline{b}$.
 We studied both the leptonic and hadronic decays of the $W$, the latter
being affected by large backgrounds. In table~\ref{tab:low} we show the
number of events (signal and
 background) expected in 15 fb$^{-1}$ for the $WH$ (leptonic case). The
sample is triggered on a
 high $P_T$ lepton and is selected on additional requirements on missing
$E_T$ ($>$20 GeV)
 and two b-tagged jets. 
\begin{table}[htb]
\raggedright
{
\begin{tabular}{|l|ccc|}
\hline
$\bf{M_H}$ &110&120&130\\
\hline
Signal events&75&60&45\\
&&&\\
Wb$\overline{b}$&435&375&285\\
$WZ$&90&60&30\\
$t\overline{t}$&225&300&330\\
single top&105&135&135\\
\hline
$\bf{S/\sqrt{B}}$&2.6&2.0&1.6\\
\hline
\end{tabular}
}
\raggedleft
\begin{tabular}{|l|ccc|}
\hline
$\bf{M_H}$ &110&120&130\\
\hline
Signal events&69&48&31.5\\
Zb$\overline{b}$&84&69&52.5\\
Wb$\overline{b}$&100&81&63\\
$ZZ$&43.5&3&0.0\\
$t\overline{t}$&70.5&64.5&52.5\\
single top&79.5&70.5&57\\
\hline
$\bf{S/\sqrt{B}}$&2.4&2.0&1.5\\
\hline
\end{tabular}

\caption{Signal and background in 15fb$^{-1}$ in the low mass
region, $WH$ channel (left), $ZH$ channel (right)} 
\label{tab:low}
\end{table} 
To enhance sensitivity to identify $b$ jets from $H$ decay, Run I studies shown that
 optimal results are obtained by a combination of a "strict" b-tagging 
algorithm
 (SECVTX) and of a "loose" algorithm. In this way the mis-tagging per
jet is kept
at $\approx$ 1 \% leaving a situation dominated by physics background.
To reduce background from $t\overline{t}$, all events with a second
lepton are rejected, as well as events with additional jets.
The final backgrounds are 
$Wb\overline{b}$, $t\overline{t}$, $WZ$ and
 single top. All of them, but for the single top, were measured in Run I
and will be done again with better statistics in Run IIa.

Similar backgrounds plague the $ZH$ channel. The most promising decay of
the Z is the
 $Z\rightarrow \nu\nu$ channel where the events are collected by a missing
$E_T$($>$ 35 GeV)trigger.    
The selection
requires two b-tagged (loose and tight) jets, distance between the
missing $E_T$ vector and the closest jet to be $>$0.5 in $\eta -\varphi$
space, and the sum of
 hadronic energy to be below 175 GeV (to reduce $t\overline{t}$
background)(table~\ref{tab:low}). 
Less encouraging results are obtained in the full hadronic channel, where
the $W$ associated to the Higgs decays in two jets.
$S/\sqrt(B)$ ratio is
about or below 0.2 in 15 fb$^{-1}$ for low mass Higgs. 
In the "high mass" region , where the WW* decay channel opens, the
$gg\rightarrow H$
mechanism becomes the most important source due to the low level of background.
While the main
focus is on the final state $ll\nu\nu$, the trilepton channel where
W(Z)H,$H\rightarrow W(Z)W^*W^*$, provides a sizeable contribution. In 15
fb$^{-1}$ we
 expect S/$\sqrt{B}$ in excess of 2 for $M_{H} > $150 GeV/c$^2$.
\begin{table}[htb]
\centering
\begin{tabular}{|l|ccccc|}
\hline
$\bf{M_H}$ &140&150&160&170&180\\
\hline
Signal events&39&42&22.5&16.5&15\\
Total background&660&450&66&36&57\\
\hline
$\bf{S/\sqrt{B}}$&1.5&2.0&2.8&2.75&2.0\\
\hline
\end{tabular}
\caption{Signal and background in 15fb$^{-1}$in the high mass region}
\label{tab:high}
\end{table}   
Table~\ref{tab:high} shows the results relative to this region.
\begin{figure}[htb]
\centering
\epsfig{figure=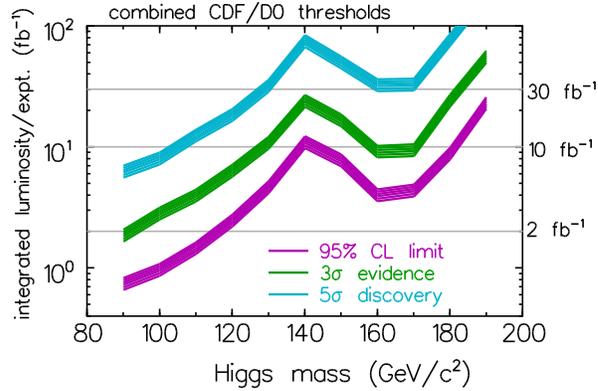,width=.5\linewidth}
\caption{Final results for Higgs searches, CDF and D0 combined}
\label{fig:final}
\end{figure}  
The (combined) CDF and D0 expectations are shown in fig.\ref{fig:final} 
where
$M_{H}$ vs. luminosity is shown~\cite{higgswg}for the whole mass range
$80<M_H<180 GeV/c^2$. 
The three bands correspond to 95 \% CL, 3$\sigma$
and 5 $\sigma$ effect. 
The lower limit of the band corresponds to the
results obtained by this study, while the width has been
obtained by considering a (positive only) 30\% uncertainty.

\section {Run IIb}
While 15 fb$^{-1}$ appear to be the amount of data needed, only 2
fb$^{-1}$
are foreseen for
Run IIa. The Fermilab Directorate launched a program to upgrade the
Tevatron
accelerator complex in order to deliver 15 fb$^{-1}$ by 2007. In order to
work the upgrade
requires the use of electron cooling to efficiently recycle the
antiprotons, as well as to
keep the beam-beam tune shift under control using the  TEL
(Tevatron Electron
Lensing) which will allow a relatively modest crossing angle. 
Therefore, although upgrade of the
Tevatron seems feasible, there are some challenges to be met in order to
bring the
 instantaneous luminosity of the machine from $1\div 2$x10$^{32}
cm^{-2}s^{-1}$ to 5x10$^{32}$, i.e. to about 5 interactions/crossing.
At the same time there will be a challenge for the detectors to match this new
environment.
The replacement of (at least) the innermost layers of silicon
detectors is already foreseen as they were designed for a Run II of 2
fb$^{-1}$
while the detectors must survive to 7 times as much luminosity.
An aggressive R\&D and design phase has started to define the other
modifications which are needed to match this new challenge.
\section{Conclusion and Acknowledgments}
The CDF and D0 experiments were able to set the tools for Higgs searches
at the Tevatron Collider already in Run I. While the luminosity foreseen
for Run IIa is probably not enough to discover the Higgs, the Run
IIb, which will allow each detector to collect 15fb${^-1}$. This, together
with the improvements of the tracking and calorimeters of each detector,
will open the possibility to discover the Higgs both in the low mass
(below 130 GeV/c$^2$) and above 150 GeV/c$^2$.
I would like to thanks E.Barberis, W.Yao, G.Velev from CDF and D0 for 
the discussions and comments and P.Derwent of the Fermilab Beams Divisions
for the information on the Run IIb upgrade projects of the accelerator.

%
\end{document}